# Aging effect in the BESIII drift chamber

DONG Ming-Yi(董明义)[1,2] XIU Qing-Lei(修青磊)[1,2] WU Ling-Hui(伍灵慧)[1] WU Zhi(吴智)[1,2]
QIN Zhong-Hua(秦中华)[1,2] SHEN Pin (沈品)[1] AN Fen-Fen(安芬芬)[1] JU Xu-Dong(鞠旭东)[1,2,3]
LIU Yi(刘义)[1,2,3] ZHU Kai (朱凯)[1] OU-YANG Qun(欧阳群)[1,2] CHEN Yuan-Bo(陈元柏)[1,2]

[1] Institute of High Energy Physics, CAS, Beijing 100049, China
[2] State Key Laboratory of Particle Detection and Electronics, Beijing 100049, China
[3] University of Chinese Academy of Sciences, Beijing 100049, China

**Abstract:** As the main tracking detector of BESIII, the drift chamber provides accurate measurements of the position and the momentum of the charged particles produced in $e^+e^-$ collisions at BEPCII. After six years of operation, the drift chamber is suffering from aging problems due to huge beam related background. The gains of the cells in the first ten layers show an obvious decrease, reaching a maximum decrease of about 29% for the first layer cells. Two calculation methods for the gain change (Bhabha events and accumulated charges with 0.3% aging ratio for inner chamber cells) give almost the same results. For the Malter effect encountered by the inner drift chamber in January 2012, about 0.2% water vapor was added to the MDC gas mixture to solve this cathode aging problem. These results provide an important reference for MDC operating high voltage settings and the upgrade of the inner drift chamber.

## 1. Introduction

The drift chamber (MDC) is the main tracking detector of the Beijing Spectrometer (BESIII), used for accurate measurements of the position and momentum of charged particles produced in $e^+e^-$ collisions [1]. The MDC is a cylindrical chamber consisting of an inner chamber including 8 layers of sense wires and an outer chamber including 35 layers of sense wires, with a helium based mixture gas (He / $C_3H_8$ = 60 : 40), working in a 1 T magnetic field. The cell dimension is about 12 mm × 12 mm for the inner chamber and 16.2 mm × 16.2 mm for the outer chamber. The average gas gain of the MDC is about $3\times10^4$ at the reference operating high voltage of about 2200 V [2] [3], but after running for six years, the gain of the MDC cells shows an obvious decrease, which leads to the SNR (signal noise ratio) becoming bad and results in a reduction of MDC performance, such as the spatial resolution and tracking efficiency. This is called the aging effect.

The aging effect includes anode aging and cathode aging. For anode aging, the gas polymer, being attracted by the high electric field near the sense wire, condenses on the wire surface as various forms: thin films, whiskers and powder. These deposits can be insulating or electrically conductive, which usually cause a gain loss due to the increase in the effective diameter of the sense wire. An additional gain drop occurs due to the reduction of the electric field caused by an accumulation of charges on an insulating layer. The variation of the deposit thickness along the wire causes a variation of the gain and, consequently, a loss in pulse-height resolution. For cathode aging, a polymer formation deposits on the cathode surfaces. This insulating layer prevents the neutralization of positive ions, leading to the formation of a surface charge. The charge induces a high electric field which can be enhanced enough to extract electrons from the cathode. Most of them recombine with positive ions immediately, but some of them drift to the anode and generate avalanches at the sense wire. The avalanche positive ions come back to the cathode, enhance the electric field of the insulating layer, and thus feed a continuous, self-sustaining local discharge in the chamber without external irradiation. This effect is called the Malter effect [4].

Aging studies are one of the main subjects in the construction and the whole working period of the drift chamber. In this article, we discuss the anode aging study of the MDC, and then introduce the Malter effect encountered by the MDC in 2012.

## 2. Estimation of MDC hit rate

In a certain working time, the accumulated charges of MDC cells, which will contribute to the aging effect, depend on the hit rate. Furthermore, the hit rate also determines the occupancy of the MDC, which will have an effect on the tracking efficiency. Experimental investigations show that the hit rate of the MDC mostly comes from beam related backgrounds, including synchrotron radiation photons, the beam-gas and Touschek [5] lost particles, which are sensitive to the beam position, the beam cross angle, the beam pipe vacuum close to the interaction point, and the apertures of the movable beam collimators in $e^+$ and $e^-$ rings [6]. We can reduce the beam related backgrounds by optimizing the machine parameters mentioned above.

Because the beam related backgrounds decrease with the beam current reduction after beam injection, in order to estimate the maximum hit rate, the data used for MDC hit rate calculation are from the random trigger events at the beginning of a run. The hit rate $H_r$ can be described by Eq. (1).

$$H_r = \frac{N}{T_r \cdot \Delta t \cdot S_{\Delta t}} \quad (1)$$

where $N$ is the total number of hits of a single cell in a very short time interval of $\Delta t$, for example 30 s, in which we can suppose that the hit rate is constant; $T_r = 60$ Hz is the random trigger rate in the data taking period; and $S_{\Delta t} = 2$ μs is the sampling time window of MDC time measurement for each trigger. Fig. 1 shows the hit rate per wire, as a function of MDC layer, in each year. The single wire hit rate reduces dramatically from the inner layers to the outer ones of the MDC. The maximum hit rate of the innermost layer was 170 kHz per sense wire in 2010, but was reduced by more than half in the last two years by optimizing the machine parameters through experiments. We can also obtain the occupancy according to the hit rate and the cell number of each layer, as shown in Fig. 2. The maximum of occupancy of the first layer was about 34% in 2010, which led to a low tracking efficiency.

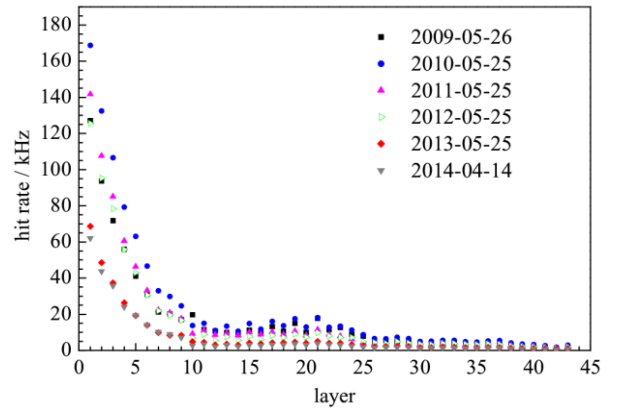

Fig. 1 Single wire hit rate each year as a function of MDC layer

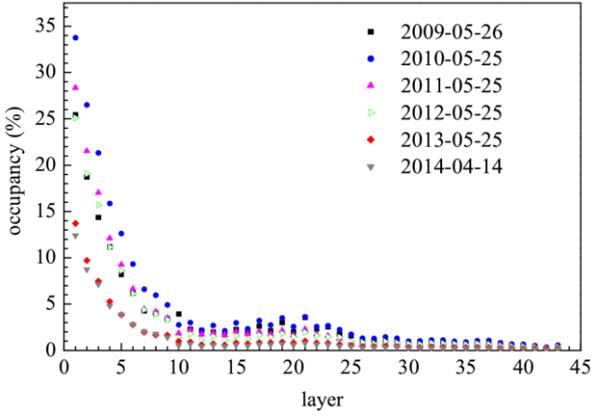

Fig. 2 MDC occupancy each year as a function of layer

## 3. MDC gain decrease

The decrease of MDC gain has been analyzed using Bhabha events in which the electrons and positrons are nearly minimum ionizing particles and they lose nearly the same energy in a cell. We reconstruct the tracks of the Bhabha events, and use the hit information of each layer in the track to get the charge spectrum of the minimum ionizing particles. The peak value of the charge distribution is obtained through a Gaussian fit with a truncation to minimize the error from the Landau tail. From the change of the charge peak value in each year, we can get the decrease of the effective gas gain of the MDC cells.

The impacts of MDC operating gas temperature and pressure changes on the gas gain are taken into account in the analysis, and are corrected by the Diethorn equation [7]. In order to reduce the error from these gas factors, we select data samples taken from almost the same time (the end of May) in each year, in which the gas pressure and temperature only have small changes.

Because data taking with the MDC started in 2009, we assume that the gas gain of the MDC at the beginning of data taking in 2009 is an initial value. The gain decrease can be obtained by comparing the gain in each year with the one in 2009, as shown in Fig. 3. The gains of the cells in the first 10 layers experience an obvious decrease after 6 years of operation, reaching a maximum of about 29% for the first layer, while the gains of the cells in the outer layers far away from the interaction point have almost no change.

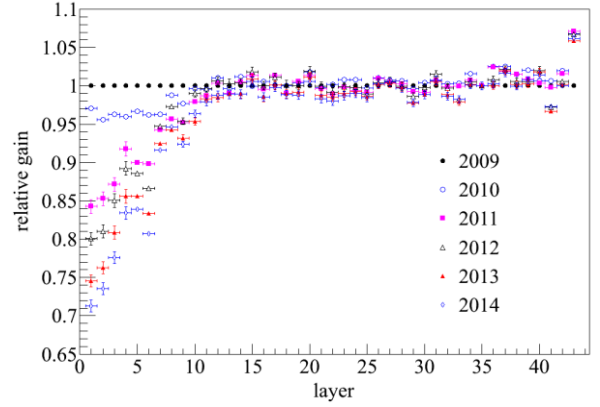

Fig. 3 Gain decrease of the cells in each year as a function of MDC layer from Bhabha events

## 4. Cell accumulated charge

The aging effect depends on the total radiation dose, which is roughly proportional to the accumulated charge of a cell, so the gain change of a cell can also be extracted from the accumulated charge and aging ratio.

The aging ratio, $R$, is usually defined as a normalized derivative of the gas gain with respect to the accumulated charge per unit length, and can be described by Eq. (2).

$$R = -\frac{1}{Q} \cdot \frac{dG}{G_0} \, \% / (mC/cm) \qquad (2)$$

where $G_0$ is the initial gain, $G$ is the current value of the gain and $Q$ is the accumulated charges of a cell per sense wire length. The aging ratio depends on the cell structure, high voltage and operating gas. For a given working condition, we can assume that the aging ratio of the MDC is a constant value. The aging ratio has already been studied during the construction of the MDC using a prototype with the same cell structure, high voltage settings and operating gas mixture as the MDC. The aging ratio is about 0.3% for the inner chamber cells and about 0.2% for the outer chamber cells [8].

The accumulated charges of MDC sense wires in the last 6 years have been calculated by integrating the currents of the cells in each year, as shown in Fig. 4. The accumulated charges of the innermost sense wires are about 105 mC / cm and then the charges reduce gradually for the sense wires far away from the interaction point. According to the aging ratio and accumulated charges, we can also get the gain change of MDC in each year, as shown in Fig. 5, which is consistent with the result calculated from Bhabha events. The gains of the cells in the first 10 layers decreased quickly before 2012, but kept a maximum decrease of about 4% for the first layer cells in the last two years, due to a relatively low background.

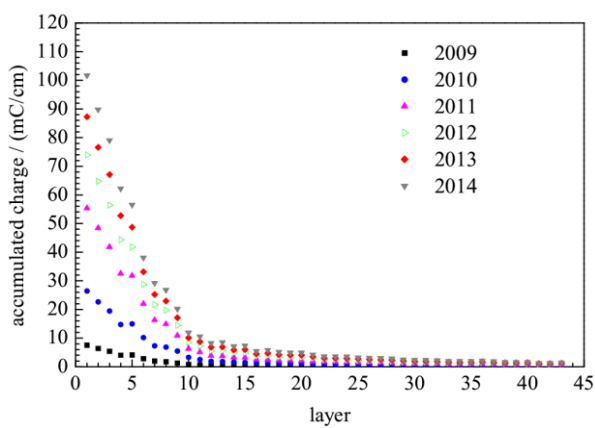

Fig. 4 Accumulated charge of the cells as a function of MDC layer in each year

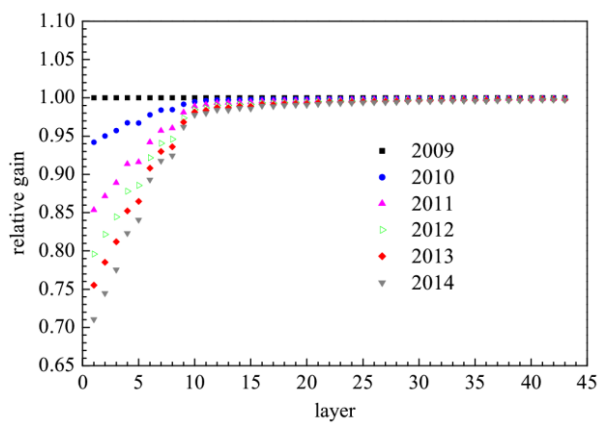

Fig. 5 Gain decrease of the cells as a function of MDC layer in each year, from accumulated charge and aging ratio

## 5. MDC cathode aging

The MDC encountered the Malter effect in January 2012, when the currents of some cells in the inner chamber jumped to a high level during data taking without high voltage change and beam lost. The increasing current, which reached a maximum of a few microamperes, did not disappear even after stopping the beam irradiation, until the high voltage was powered off, as shown in Fig. 6. It was verified to be the Malter effect, a kind of cathode aging, which can easily occur in gas chambers with high gas gain working at huge background. The Malter effect has been encountered by several international high energy experiments; for example, the BaBar drift chamber encountered the Malter effect during its commissioning [9]. For the BES experiments, however, it was the first time to observe this kind of discharge. Because neighboring cells of the MDC share the same field wires, Malter discharge can spread fast in the inner chamber, which leads to more and more affected cells that cannot work normally due to the high current.

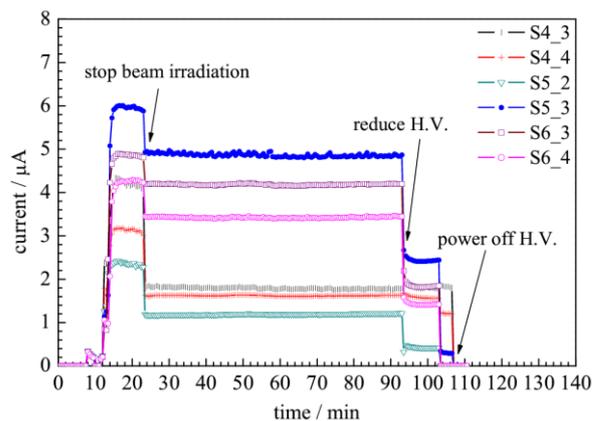

Fig. 6 The current of some cells as a function of time, reaching a maximum of 6 μA due to Malter discharge. The current did not disappear even after stopping beam irradiation, until the high voltage was powered off.

Firstly, we divided the high voltage system of the MDC inner chamber into several sectors, and implemented independent monitoring of the cell currents in a sector which include 8 cells, trying to find the affected cells, and then turned off the high voltage of the affected cells in time to prevent the spread. A drift chamber prototype system was set up for gas additive experiments, and 5% $CO_2$ or 0.1%-0.35% water vapor respectively was added to the operating gas of the prototype respectively for testing. Based on the test results, we added 5% $CO_2$ to the MDC operating gas, and trained the MDC for about a week with electron beam current increasing slowly. From February 27 to March 30, the MDC worked normally with 5% $CO_2$ at the colliding beam current less than 500 mA without Malter discharge, but the gas gain of the MDC decreased by about 23% with the 5% $CO_2$ additive. In addition, Malter discharge was induced once by high beam related background at the very beginning of this run period. A water vapor system was therefore developed, and from April 1, about 0.2% water vapor, which replaced the 5% $CO_2$, was added to the MDC operating gas. The MDC worked stably with a gas gain decrease of about 9%, and no Malter discharge has been observed since then.

## 6. Discussion and conclusion

We have studied anode aging in the BESIII MDC, which results in a decrease of the gains of MDC cells, both by Bhabha events and accumulated charges, with a 0.3% aging ratio for the inner chamber cells. The results of the two calculation methods are in agreement with each other. After six years of operation, the gains of the cells in the first ten layers show an obvious decrease, reaching a maximum decrease of about 29% for the first layer cells. The gain decrease stayed at a maximum decrease of about 4% for the first layer cells in the last two years due to a relatively low background. The gains of the cells can be increased to compensate for the aging effect by increasing the operating high voltage, but the high voltage increase means that more charge will accumulate on the sense wires, which will accelerate MDC aging.

For the Malter effect encountered by the inner drift chamber in January 2012, about 0.2% water vapor was added to the MDC gas mixture to solve this cathode aging problem.

The aging study results provide an important reference for MDC operating high voltage settings and the upgrade of the inner drift chamber [10].

*This work is supported in part by the CAS Center for Excellence in Particle Physics (CCEPP).*

______________________________________